\documentstyle[twocolumn,prb,aps,psfig,epsf,fleqn,ifthen]{revtex}  

\begin{document}  
  
  
\vskip 1 cm  
\title{A few electrons per ion scenario for the $B=0$ metal-insulator transition 
in two dimensions}  
\author{T.M. Klapwijk \thanks{Correspondence should be 
addressed to T.M.K (e-mail: klapwijk@phys.rug.nl) }}  
\address{Center for Quantized Electronic Structures (QUEST) and Institute for 
Theoretical Physics, University of California at Santa 
Barbara, California 93106-4060 \\ and}
\address{Department of Applied Physics and Materials Science Centre,   
University of Groningen,  
Nijenborgh 4, 9747 AG Groningen, The Netherlands}     
\author{S. Das Sarma}
\address{Institute for Theoretical Physics, University of California at Santa Barbara, 
California 93106-4060 \\ and}
\address{Department of Physics, University of Maryland, College Park, Maryland 
20742-4111} 
                                                  
\date{\today}  
\vskip 0.5 cm 
\maketitle  
  
\begin{abstract}                                               
We argue on the basis of experimental numbers that the $B=0$ metal-insulator transition in 
two dimensions, observed in Si-MOSFETs and in other two-dimensional systems, is likely 
to be due to a few strongly interacting electrons, which also interact strongly with the 
random positively ionized impurities. At the insulating side the electrons are all bound 
in pairs to the ions. On the metallic side free electrons exist which are scattered 
by ions dressed with electron-pairs and therefore alter the bare scattering potential 
of the ions. The physics at the metallic side of the transition is argued to be 
controlled by the classical to quantum transport cross-over leading to the observed 
non-monotonous dependence of the resistivity on temperature. This few electrons per ion 
scenario appears to be an experimentally realistic and testable scenario, which can also 
serve as a starting point for further theoretical analysis of the two-dimensional 
metal-insulator transition.       
\end{abstract}  
  
  
\subsection*{Introduction}

A number of recent\cite{KravPRB94,KravPRB95,KravPRL96,HaneinPRL98,SimmonsPRL98} striking\cite{PhysToday} experimental observations 
on the low temperature ($0.1-10  
\, K$) transport properties of low density two dimensional electron/hole systems (2DES) 
have been interpreted on the basis of 2D metal-insulator transition\cite{PhysToday} (2D M-I-T), which is 
nominally prohibited by weak localization theories\cite{Abrahams79}. A critical density $n_c$ ($\leq 
\sim 10^{11} cm^{-2}$) for Si-MOSFETs separates the "metallic" and the "insulating" sides 
of the transition with the resistivity showing a strong exponential (positive) 
temperature dependence in the insulating side and a weak negative temperature dependence 
(at least at lower temperatures) in the metallic side. Some of the published data show a 
remarkable scaling behavior in carrier density ($n_s$) and temperature 
($T$), which is consistent with a M-I-T $T=0$ quantum phase transition 
phenomenology\cite{Dobrosavljevic,Sondhi97}, 
and therefore a great deal of theoretical attention has been focused on this "2D M-I-T" 
phenomenon, with theoretical proposals\cite{Castellani98,Phillips98,Belitz98,HeXie,Chakravarty98,Si98}  
 ranging from the phenomenon being a 
superconductor-to-insulator transition to non-Fermi liquid theories. In addition, a more 
conventional explanation has been proposed by Alsthuler and Maslov\cite{Altshuler98} 
based on charge traps in the silicon-dioxide.  
 
In this article we provide a critical phenomenological discussion of the experimental 
system, concentrating almost entirely on the $Si/SiO_2$ MOSFET systems used by Kravchenko 
and collaborators\cite{KravPRB94,KravPRB95,KravPRL96} in their pioneering work. For this system a large body of detailed 
experimental published data is available, and the most compelling evidence for the 
existence of a 2D M-I-T has been established in this system. We believe that a detailed 
qualitative (perhaps even a semi-quantitative) understanding of the temperature dependent 
resistivity, $\rho(T)$, around the critical density can be developed for Si-MOSFETs on 
the basis of the considerations proposed here,  in particular by focusing on the 
experimental parameters of Si-MOSFETs. Our physical considerations described below 
should serve to provide rather stringent experimental constraints for possible theories 
of the 2D M-I-T. We also provide a physically motivated conjecture about the mechanism 
that drives the transition in these systems. Our proposed 2D M-I-T scenario is precise 
and directly experimentally testable.

\subsection*{Mobility as a function of electron density}

The mobility of MOSFETs, which can be defined only on the metallic side of the 
transition, $n_s>n_c$, is limited by scattering from surface roughness and ionized impurities\cite{Ando}. In 
high-mobility samples\cite{Kruithof91a,Kruithof91b,Heemskerk98,Pudalov93,Vyrodov87}, such as used to measure the metal-insulator transition, high 
resistivity Si material is used, typical resistivities in excess of $100 \Omega\, cm$ or 
the density of donors less than $10^{14} cm^{-3}$, and therefore scattering by 
ionized donors can be ignored. The carrier density is varied by applying a voltage to 
the gate electrode which is separated from the silicon by insulating $SiO_2$. At the interface a 
triangular potential well, "the inversion layer", is formed of which only the lowest 
subband is occupied at low temperatures. 
Experimentally, one observes studying conduction-processes parallel to the $Si/SiO_2$ 
interface, a non-monotonic dependence in the low temperature 
mobility as a function of density: for decreasing density an 
increase in mobility is seen reaching a maximum, called the peak-mobility, followed by a 
rapid decrease, as the carrier density is further reduced  below $n_{max}$ where the 
mobility reaches its maximum, $\mu_{max}$.

It is understood that the decreasing and the increasing part of the mobility reflects
scattering by different physical processes. Starting at high electron densities, high applied gate 
voltages, the carriers are 
pulled towards the $Si/SiO_2$ interface and experience therefore strongly the 
surface roughness. For high mobility MOSFETs such as used in these 
experiments\cite{Kruithof91a}, the 
amplitude of the roughness is approximately $0.3 \, nm$ and the correlation length in 
the order of $7\, nm$. Upon lowering of the gate voltage the 2D gas moves away from the 
interface 
and the mobility increases. On the other hand the random ionized impurities, invariably 
present at the $Si/SiO_2$ interface as a result of the fabrication process, become more and more effective at lower electron densities, 
because the electronic screening is reduced. A high peak mobility and the associated low 
values of $n_{max}$ at which the mobility peak occurs is taken as a 
signature\cite{Kruithof91a,Kruithof91b,Heemskerk98,Pudalov93,Vyrodov87} of good quality MOSFETs because it signals both low surface roughness, 
which one can not do very much about, and also a low density of ionized impurities, 
which is controlled by working under very clean conditions in particular for the 
oxidation process.

The first point we make (which has not been emphasized in the existing theoretical 
literature on 2D M-I-T) is that the observed metal-insulator transition occurs at 
electron densities where the ionized impurity scattering dominates. In particular, 
the 2D M-I-T is only seen in samples 
with high $\mu_{max}$, and usually with $n_c \ll n_{max}$.    
In our own samples\cite{Heemskerk98} the metal-insulator transition occurs at carrier densities 
of about $8\ 10^{10}cm^{-2}$, far below the density $n_{max}$ at which we have 
the peak mobility, $6\ 10^{11}cm^{-2}$. Consequently, we are in a 
regime where carrier scattering (and therefore transport) is dominated by ionized 
impurities. In this situation ($n_c\ll n_{max}$) electron densities 
are sufficiently low so that the screening of these ionized impurities has become strongly weakened. 
This has two implications. First, it means that the scattering potential should not be looked 
at as the surface roughness, namely a hard-walled potential landscape due to fixed fluctuations 
like a rocky landscape. The potential scattering is one due to ions, 
predominantly positive,  located in laterally random positions, at the interface 
or slightly (only a few $\AA$) away from 
the interface inside the oxide. The second implication is that we have a system in which 
the electrons are facing random localized attractors, {\it i.e.} the charged impurity 
ions, which are poorly screened. Thus, electronic screening is an important ingredient 
in understanding the observed 2D M-I-T. To the best of our knowledge a 
$n_c\,> \, n_{max}$ has never been reported in Si MOSFETs, samples with the highest 
peak-mobility\cite{KravPRB94} $\mu_{max}\sim 7 \, 10^4 \, cm^2/Vs$ have $n_c\leq n_{max}/2$, whereas 
samples with a more modest peak-mobility of $\sim \, 2\, 10^4 \, cm^2/Vs$ typically have 
$n_c\leq n_{max}/5$. 

\subsection*{How many electrons per ion?}

 It has been emphasized by Kravchenko and co-workers\cite{KravPRB94,KravPRB95,Pudalov93} that to observe the 2D M-I-T the 
 electron-electron interaction energy $E_{e-e}$ should dominate over the Fermi-energy 
 $E_F$, which means 
 that $r_s=E_{e-e}/E_F=1/(\pi n_s)^{1/2} a_{B}^*$ should be substantially larger than 
 $1$ ($n_s$ the carrier density and $a_{B}^*$ the Bohr radius taking into account the 
 dielectric constant of the material). 
  The M-I-T in MOSFETs is observed around $r_s < \sim 10$. This argument has been used 
 to justify the absence of a 2D M-I-T in an electron gas in GaAs/AlGaAs 
 heterostructures. The lighter effective mass and the larger dielectric constant would 
 push this regime to much lower densities. Convincing experimental support for this aspect 
 of the scenario has been provided by the recent reports\cite{HaneinPRL98,SimmonsPRL98} on a similar 2D M-I-T in a 
 2-dimensional hole gas in GaAs/AlGaAs heterostructures in which the holes have a very 
 high effective mass. In these systems the transition occurs around $r_s\sim 18$. 
 Moreover, very recently the metal-insulator transition has also been  
 observed\cite{Hanein;n-GaAs98} in n-type  GaAs for $r_s\sim 6$. Therefore we believe that 
 there is convincing evidence that strong electron-electron interactions 
 play a key role.  
 
 Nevertheless, although this low $r_s (\sim 10)$ regime can easily be reached in all Si-MOSFETs, the 2D M-I-T 
 is only observed in high mobility MOSFETs\cite{KravPRB94,Popovic97}. Apparently, some 
 other materials parameters play a 
 role in addition to the electron-electron interaction. As we emphasized above a high 
 mobility means a relatively low density of ionized impurities at the $Si/SiO_2$ 
 interface. What is the density of these ionized impurities in practice? One way to 
 determine their density is by measuring the capacitance as a function of 
 voltage\cite{Ando}. Taking 
 into account the materials parameters of the metal gate, the oxide, and the silicon 
 we know 
 what the ideal C,V-curve should look like (in the absence of random charged impurities). 
 The observed shift from the theoretical values provides the density of ions in the oxide. 
 A second method\cite{Kruithof91b,Vyrodov87} is to perform an accurate  quantitative evaluation of 
 the mobility {\it vs.} density curve, using the well-established theory for the scattering 
 processes at densities around $n_{max}$. Both methods lead to a number of about $3\ 10^{10}\ cm^{-2}$ 
 for high-mobility MOSFETs as used for the 2D M-I-T experiments. There is very little room to 
 vary this number by more than a factor of 2. In fact it is roughly proportional to the 
 peak mobility $\mu_{max}$. A decrease in the charged impurity  
 density by a factor of 2 would raise the peak mobility\cite{Kruithof91b} from $25,000\ cm^2/Vs$ to 
 $50,000\ cm^2/Vs$, which is not observed. In fact in our own 
 samples \cite{Kruithof91a} we 
 infer a charged impurity density of $2.4\ 10^{10}\ cm^{-2}$, which means that the 
 average mutual separation is about $ 70\, nm$. Similar numbers are quoted by other 
 researchers, in particular by Pudalov et al\cite{Pudalov93}. (Incidentally 
 these numbers are much lower than those assumed in a recent manuscript of Altshuler and 
 Maslov\cite{Altshuler98}, which tries to explain the 2D M-I-T as due trapping of 
 electrons by ions, using a different mechanism than proposed by us.)   
 
Experimentally the 2D M-I-T is observed\cite{Heemskerk98} at a critical electron density $n_c$ 
of around $8\ 10^{10}\,cm^{-2}$. Comparing with the density of ions we arrive at the \emph{striking} result 
that at the critical density we have only 2 to 4 electrons per ion. This points to a 
natural explanation for the requirement of high mobility MOSFETs. Accepting the fact that the density should be 
low enough to have strong electron-electron interactions, if the mobility is lower by a 
factor of 3, for example below $10,000\, cm^2/Vs$, we have a 3 times higher density of ionic impurities 
and hence we will at a $r_s=8$ always have a situation of less than one electron per 
impurity ion at the interface. Also, if we now compare the range of electron 
densities at which the anomalous metallic state is observed down to below the critical 
density we observe that all the action takes place in a regime in which the number of electrons 
decreases from at most 5 electrons/ion to 2 electrons/ion. We believe that transport 
physics at such a high density (2-5 electrons per ion) of random charged impurity 
centers is fundamentally different from high density electron systems because screening 
becomes strongly non-linearly affected by ionic scattering and binding, leading to a 
sharp M-I-T at $n_c$ when effectively all the electrons become bound to or trapped at 
individual ions at low enough temperatures.  

\subsection*{Proposed scenario}
                                                                            
We propose therefore that the 2D M-I-T is a phenomenon occurring in a system of interacting 
electrons and random positive ions with a density ratio of at least 2 electrons per ion at 
the critical density $n_c$. From the experiments carried out by Simonian et 
al\cite{Simonian97} it is 
clear that the conducting (and insulating \cite{Kravprivate98}) states are sensitive to a parallel magnetic field which means that the spins play 
a crucial role. This observation points to the possible importance of spin-singlets. Taking 
these numbers into account and the possibility of spin singlets we arrive at the following 
conjecture. We suppose that we can divide the electrons in two groups: localized (bound) 
and delocalized (free) electrons. The localized electrons consist of two (or perhaps four) 
electrons per ion, bound with opposite spins at individual random ionic impurities. 
This binding is weak, as the binding 
energy\cite{Vinter82} at low electron density is a fraction of an $meV$. The delocalized electrons are 
scattered by impurity ions which are now dressed by two (or four) bound electrons which 
leads to a weaker scattering potential than the bare ion-potential. Hence starting from the higher 
electron density, metallic,  side the scenario is that we have free electrons scattered of 
hard-walled ionic potentials. Then a crossover takes place as the electron density is 
lowered, which involves two 
processes. Electrons are scattered more strongly by the ions because screening weakens 
as impurity scattering induced level broadening modifies screening at low density. In 
addition,  the number of free or unbound electrons decreases as more  electrons bind to 
the individual ions with decreasing $n_s$.  
Upon further lowering the electron density it 
becomes possible that two (or four) electrons are trapped at the ions, which alters 
drastically the ionic   
scattering potential. This feedback mechanism continues (electrons getting bound to 
ions) as the electron density is lowered (and screening is weakened) until at $n_s=n_c$ 
suddenly all the electrons bind to ions observed as the 2D M-I-T. Clearly this would 
mean that the metallic conductance at the critical density is determined by a 
few electrons per square, a 'critical resistivity' of a few times $h/{e^2}$ as 
is experimentally 
observed\cite{KravPRB94,Heemskerk98}. 

Once all the electrons bind to ions the system at $T=0$ is an insulator, but at finite 
temperature shows a finite conductivity either due to thermal unbinding or variable 
range hopping. The strong electron-electron interaction further helps  the trapping at 
individual ions by keeping the electrons apart so as to minimize their Coulomb potential 
energy at a cost of their kinetic energy. In this sense our insulating 2D system might 
be considered a strongly pinned Wigner glass\cite{Chakravarty98,Chui97}, \emph{i.e.} an electron crystal strongly modified by the random 
charged impurity centers which act as  the nucleation centers for localizing electrons. 
We believe that because of this few electrons per ion (FELPI-) scenario proposed here, it 
is more natural to think of the insulator as a system with few (2 to 4) electrons 
trapped or bound at charged impurity sites (with strong intersite Coulomb correlations), 
which is stabilized by gain in potential energy due to the ionic bonding. It is well 
known\cite{Yoon98} that in a purely 2D system without the charged impurity centers Wigner 
crystallization occurs at a much lower density ($r_s \sim 38$). The observed 2D M-I-T in 
our scenario is the transition to this few electrons per ion (FELPI) insulating state.  

In order to specify more precisely the experimental system we point out that the ions 
are located at the $Si/SiO_2$ interface, whereas the center of the 2-dimensional 
electron gas is, at the critical density $n_c$, at a distance of $7\, nm$ away from this interface. 
Therefore the problem of individual ions with 2 electrons attached to them is 
reminiscent of that of the so-called $D^-$ centers in GaAs/AlGaAs\cite{Marmorkos97}. 
These have been shown to exhibit a subtle dependence on the magnetic field 
including the so-called magnetic field induced dissociation, although the mutual interaction 
of the ion-electron system has not been studied yet. We also point out that the 
average distance between the electrons at the transition is about $33 \, nm$, whereas 
the Fermi wavelength is $120 \, nm$. Finally, we point out that extracting the 
scattering time naively from the mobility around $n_c$ leads to unrealistically small values of $0.1\, ps$, 
corresponding to a level broadening of $3\,meV$, in clear disagreement with the values 
obtained from Shubnikov-De Haas oscillations \cite{Kravprivate98} which point to a level broadening of the order $0.1 
\, meV$. The latter value corresponds quite well to taking a mean free path of the order 
of the 
average inter-ionic distance and the appropriate Fermi velocity of $3\, 10^6\, cm/s$.

We now consider what happens within the FELPI-scenario if one applies an external 
electric or magnetic field\cite{KravPRL96,Heemskerk98,Simonian97}. On the 
insulating side one would find an increase in conductance for a higher electric field in 
the usual way due to the electric field assisted hopping of the bound electrons. In this regime the application of a 
magnetic field eliminates (by lifting the spin degeneracy) the condition for pair 
binding, and we get more strongly 
bound electrons, of one per ion. This increased binding arises from the singlet-triplet 
bound state energy difference. So the application of a 
magnetic field on the insulating side leads to an increase in resistance. On the 
metallic side an electric field will strip the ions of bound electrons altering the scattering potential in 
making it a stronger scatterer and therefore an increase in resistance too. In addition, 
screening decreases (by a factor of two in 2D) due to the lifting of the spin-degeneracy 
leading to enhanced scattering and a consequent increase in resistance. The charged 
scattering centers behave somewhat like negative U-centers\cite{Anderson75} and the external magnetic 
field strongly affects the binding energies leading to the observed effects. As is 
evident from the magnetic field dependence of the $D^-$ centers in GaAs/AlGaAs, the 
actual details 
may depend on the model\cite{Marmorkos97}. 

\subsection*{Quantum to classical crossover}

We now consider as an example\cite{DasSarma98}  a direct consequence of our observation that the 2D M-I-T 
\emph{always} occurs in a regime dominated by random long range charged impurity scattering
(and 
not by the short range interface roughness scattering, which is effective for $n_s > n_{max}$ and $\gg 
n_c$), and therefore screening is important. For low electron density and 
charged-impurity scattering-dominated transport the detailed temperature dependence of 
the mobility can be theoretically calculated by combining the Boltzmann equation with 
the dielectric screening formalism including finite temperature and level broadening 
effects, due to the charged impurities. Results of such calculations are 
in reasonable qualitative agreement with the 
experimental observations on the temperature and density dependence of the measured 
mobility on the metallic side ($n_s>n_c$) of the M-I-T. We define the conductivity 
$\sigma=ne\mu$ and the resistivity $\rho=1/\sigma$, following the standard convention, 
and assume that $n_s$ is fixed by the gate voltage with all the $T-$ dependence of 
$\sigma$ or $\rho$ arising from $\mu(T)$ {\it i.e.} the scattering rate. The calculation of 
$\rho$ involves an energy averaging, which contributes quite significantly at higher 
temperatures, because the effective degeneracy or Fermi-temperature $T_F$ of the system 
is quite low at low values of the density. In particular for $n_s=10^{11} cm^{-2}$ we have 
$T_F=7\, K$, with $T_F\propto n_s$ in 2D systems.The effective $T_F$ may in fact be much 
lower because only the unbound electrons contribute to the Fermi degeneracy. The asymptotic $\rho(T)$ arising from 
charged impurity scattering at effective low $T<T_F$ and high temperatures $T>T_F$ can 
be calculated and one gets: $ \rho(T)\sim\rho_0+AT\equiv \rho_q$ for $T<T_F$ and 
$\rho(T)\sim BT^{-1}\equiv \rho_c$ for $T>T_F$. In calculating these asymptotic forms, we take into account only the 
screened charged impurity scattering, and all the temperature dependence arises from 
thermal smearing of the Fermi distribution function in screening and thermal averaging. 
All other sources of temperature dependence, such as phonons, weak localization, and 
interaction corrections to scattering have been neglected as quantitatively umimportant 
for Si MOSFETs in the temperature range ($T\sim 0.1 - 10 \,K$) of interest. The 
coefficients $A$ and $B$ arise from screening and energy-averaging corrections, respectively. 
The asymptotic low ($\rho_q$)  and high ($\rho_c$) temperature resistivities given above 
correspond to the quantum ($T\ll T_F$) and the classical ($T \gg T_F$) carrier diffusion 
transport regimes respectively. The truly amazing thing to note about Si MOSFETs 
exhibiting a 2D M-I-T is that $n_c$ is so low,  $\sim 10^{11} cm^{-2}$, and hence 
$T_F<\sim 7\, K$ that the 'metallic' system just above the M-I-T shows a classical to quantum transport crossover in the narrow temperature range of 
$T\sim 0.1 - 10\, K$. This transport crossover is reflected in the 
observed\cite{KravPRB94,HaneinPRL98} strong 
non-monotonicity of $\rho(T)$ on the metallic side of the transition ($n_s>n_c$), where 
$\rho(T)$ invariably rises with temperature for higher temperatures (as in $\rho_c$) 
before exhibiting the metallic behavior of a positive temperature coefficient (as in 
$\rho_q$) at lower temperatures where $\rho(T)$ decreases with lowering temperatures.   
                                                                     
Detailed calculations show that at very low temperatures, when $T\ll T_D$ with 
$T_D\sim\Gamma/k_B$ as the Dingle temperature of the system and $\Gamma$ the 
impurity scattering induced quantum level broadening, the coefficient $A\propto T$ and 
$\rho(T)\sim \rho_0 + O(T^2)$ at the lowest temperatures, becoming $\rho(T)\sim 
\rho_0+O(T)$ for $T>T_D$. The impurity scattering limited $\rho(T)$ 
thus shows a crossover (for $n_s>n_c$) in its temperature depedence with $\rho(T)$ being 
'metallic' for $T<T^*$ and 'insulating' for $T>T^*$ with $T^*\sim T_F/3$, being 
approximately the crossover temperature. Thus for higher 
densities $T^*$ moves to higher temperatures, and the 'insulating' behavior on the 
metallic side gets 
progressively suppressed as phonons become more important. We emphasize that the 
temperature dependence of $\rho(T)$ arising from these effects could be quite large (a 
factor of 3 to 4 is usual on the metallic side, and a factor of 10 is quite possible) at 
these low densities. 

The observed strong non-monotonicity in the experimental $\rho(T)$ for $n_s\sim n_c$, 
which is naturally explained in our theory involving only charged impurity scattering, 
provides further qualitative and quantitative support  for the FELPI scenario proposed in 
this article. Additional temperature dependence may also arise at low temperatures on 
the metallic side as the temperature becomes comparable to electronic binding energies 
(to ions), reducing screening dramatically. 

\subsection*{Quantum phase transition?}

The 2D M-I-T in Si MOSFETs and other 2D systems have been interpreted on the basis of a 
$T=0$ quantum phase transition from a metallic phase (for $n_s\geq n_c$) to an 
insulating phase (for $n_s\leq n_c$) driven by the carrier density. The main support for 
the quantum phase transition scenario comes from the scaling collapse of $\rho(T,\delta 
n_s=|n_s-n_c|$) data in Si MOSFETs around the transition point $n_s=n_c$ at low 
temperatures\cite{KravPRB94,KravPRB95,KravPRL96}. This interpretation of the 2D M-I-T as a quantum phase transition has 
understandably attracted a great deal of theoretical attention\cite{PhysToday}, particularly because 
the non-interacting non-linear sigma model based weak localization or single parameter 
scaling theories\cite{Dobrosavljevic} rule out such a phase transition in 2D as all states 
are at least 
weakly localized (in orthogonal and unitary ensembles -- we will ignore the symplectic 
case relevant for spin-orbit scattering disorder), and any 2D M-I-T, in this widely 
accepted standard theory, can only be a crossover from a weak localization to a strong 
localization regime as $n_s$ decreases. We discuss the quantum phase transition aspects 
 of the 2D M-I-T in light of our proposed FELPI-scenario in this section. 
 
 In our proposed scenario the important issue we address is the existence of the sharp 
 density $n_c$ separating apparent strongly metallic ($n_s>n_c$) and strongly insulating 
 ($n_s<n_c$) low temperature $\rho(T)$ behavior. We provide a specific microscopic 
 mechanism (FELPI) where the fact that the 2D M-I-T always occurs in a situation 
 involving few electrons per random charged impurity ion center plays a crucial role: 
 for $n_s<n_c$, all the electrons are tightly bound or trapped in pairs to charged ions 
 at the interface (without any 'free' electrons available for metallic transport), 
 creating a strongly insulating state where only activated or variable range hopping 
 transport is possible at low temperatures; for $n_s>n_c$, there are free electrons 
 weakly scattering from negative U-centers\cite{Anderson75} created by charged ions with electron pairs 
 bound to them, creating a strongly metallic state at low temperatures because 
 scattering is substantially reduced by the bound electron pairs screening the ion. We 
 can qualitatively explain the strong non-monotonic $\rho(T)$ on the metallic side 
 ($n_s>n_c$) as an interplay of screening and thermal averaging as the system makes a 
 quantum to classical transport crossover for $T\sim 0.1-10\,K$ with $T_F\sim 5\,K$. We 
 can also explain the strong magnetic field dependence based on the drastic decrease of 
 screening on the metallic side and a strong increase in the binding energy on the 
 insulating side due to the singlet-triplet transition. Electric field effects are 
 easily explained in our scenario as field-induced stripping or tunneling of electrons 
 due to their relatively small binding energies. Crucial factors in our FELPI-scenario 
 are the dominance of charged impurity scattering controlling the 2D M-I-T physics, the 
 small ($\sim 0.1 - 0.5 meV$) electronic binding energies to charged ions at the 
 interface, the temperature dependence of the electronic screening, the low values of 
 the degeneracy temperature causing the quantum to classical crossover, and the 
 nonlinear self-consistent interplay of scattering/binding/screening between electrons 
 and ions. All of these factors arise because of the low electron density in the system, 
 which also produces strong inter-electron Coulomb correlations further 
 enhancing the strongly insulating state, which takes on the character of a very strongly 
 disordered electron solid (where few electrons are pinned or bound to individual random 
 charged ions) or equivalently a strongly pinned Wigner glass.
 
 It should be obvious from the above summary of the FELPI-scenario that the proposed 
 microscopic mechanism makes no reference to a quantum phase transition, and in fact, 
 based on our currently existing analysis of the problem, we cannot rule out a rapid 
 crossover at $n_s\sim n_c$ from a metallic-like phase (for $n_s>n_c$) to a strong 
 insulator phase (for $n_s<n_c$) as the number of electrons per ion in our scenario 
 decreases from being above 2 (or, some other such small number) in the metallic phase 
 to being below 2 in the insulating phase -- by definition such a crossover in our 
 FELPI-scenario will occur at a sharp electron density $n_s=n_c$ because the "M-I-T" 
 takes place precisely as the number of electrons per charged ion passes through a fixed 
 small number (taken to be 2 for our discussion, but it could be 4 with no loss of 
 generality). Within our scenario the low density ($n_s<n_c$) phase is a strongly 
 localized insulator by definition because all the carriers are strongly bound or 
 trapped at random charged ion centers. The 'metallic' phase ($n_s>n_c$) in our 
 scenario, however, may very well be a weakly localized (rather than truly extended) 
 phase at $T=0$ -- our theoretical mechanism is completely insensitive to the eventual 
 $T=0$ phase of the 'metallic' electrons for $n_s>n_c$. What is important is that the 
 system behaves as an 'effective metal' at the lowest temperatures typically attainable 
 ($T>0.1\,K$) in the 2D M-I-T experiments the crossover to weakly localized 
 behavior could occur at substantially lower temperatures; in addition the temperature 
 dependence arising from screening, as considered in our work, is roughly two orders of 
 magnitude stronger than weak localization induced temperature corrections in the 
 experimentally relevant $0.1-10\, K$ range, again making the $T=0$ fate of the metallic 
 phase inconsequential for understanding the M-I-T experiments. 
 
 While our FELPI scenario is a quantum scenario (because it involves quantum binding of 
 few electrons per ion at low temperatures) we do not require any quantum phase transition 
 to qualitatively understand the 2D M-I-T because our microscopic mechanism specifically 
 incorporates a sharp density ($n_s=n_c$) at which the system changes from an 'effective 
 metal' to a strong insulator. The transition is obviously continuous in our picture 
 because it is induced by a continuous change in the electron density. We cannot rule 
 out a quantum phase transition within our model, and whether there is a quantum phase 
 transition driving the 2D M-I-T or not is beyond the scope of our paper, and is in fact 
 quite irrelevant, in our opinion, to understanding the experimental data because the 
 data are consistent with the metallic phase being a weakly localized system where the 
 weak localization physics would only manifest itself at experimentally inaccessible low 
 temperatures. 
 
 In this context it is important to point out that while the 2D M-I-T itself seems to be 
 a generic phenomenon in the sense that the basic transition has now been observed in 
 several different classes of 2D systems, the scaling collapse of the $\rho(T,\delta 
n_s=|n_s-n_c|)$ data has only been reported for Si MOSFETs. Even in the Si-MOSFETs, 
where the scaling behavior of the 2D M-I-T has been reported, one can legitimately 
question the empirical details of the scaling collapse in the sense that 
$\rho(T,n_s>n_c)$ shows considerable non-monotonicity as a function of temperature in 
the $T=0.1-4\, K$ range (which is naturally explained in our theory as described above) 
and therefore, by definition, it is impossible to have true scaling behavior except in 
a very small temperature and density window, where the concept of scaling becomes not 
particularly meaningful. Our view is that the essence of the 2D M-I-T can be understood 
on the basis of the FELPI scenario without invoking or requiring any quantum phase 
transition, and the existence (or not) of a true quantum phase transition in the 
2D M-I-T phenomenon is open.

\subsection*{2 dimensional hole gas} 

Because of the experimental similarity the same physics must play a role in other 
reported 2D M-I-T {\it e.g.} two-dimensional 
hole gases. We argue that the FELPI-scenario plays a role 
in these systems too.  The high mobility obtained in MBE-grown samples is a result of eliminating 
interface roughness and eliminating ionized donors in the conducting plane. Scattering is 
limited to the so-called remote ionized impurities. In the sample studied by Hanein et 
al\cite{HaneinPRL98} the active 
layer is in a $150\ nm$ thick quantum well capped by a p-doped GaAs top layer. Unfortunately 
the density of dopants is not given. It is clear though that the range over which the carrier 
density is varied runs from $0.089\ 10^{11}\ cm^{-2}$ to  $0.64\ 10^{11}\ cm^{-2}$ with a 
critical value located somewhere around $0.25\ 10^{11}\ cm^{-2}$. If we would assume that 
this is also at roughly 3 holes per ion, the total range is from 1 to 7 holes per ion, 
which is consistent with the FELPI scenario. 

In the sample studied by Simmons et al\cite{SimmonsPRL98} a 20 nm thick quantum well is used 
and the nearby GaAs layer is Si modulation doped. Also in this case the density of dopants is not 
mentioned. In this experiment one finds a critical value of $n_c\sim 0.5\ 10^{10}\ cm^{-2}$. If we 
again assume that the critical value is at about 3 holes per ion in this experiment the total 
change occurs in a range from $0.3-0.7\ 10^{11}\ cm^{-2}$ or from 2 to 4 holes per ion.

While the actual numbers and the details are not as well known in the other systems as in Si MOSFETs, we 
can, however, assert that the FELPI scenario predicts a much lower value for $n_c$ 
because the charged impurities are further away from the electrons (holes). The same 
non-monotonicity in $\rho(T)$, as discussed here, is seen\cite{HaneinPRL98} in these 
GaAs systems also. In addition the recently observed\cite{Hamilton98} re-entrant 
insulator-metal transition at higher hole-densities might easily fit into 
our qualitative picture as evidencing the onset of binding of holes to the ions, and the 
re-entrance also is evidence in favor of crossover physics.
 
\subsection*{Conclusions}

We have proposed a theoretical scenario, based on the few-electrons-per-ion (FELPI) 
mechanism, for the observed 2D M-I-T in Si MOSFETs. Our theoretical scenario, in 
contrast to most other proposed theories of the phenomenon, makes extensive use of the 
actual experimental parameter values operating in the phenomenon -- in particular we 
believe that the crucial process is the dominance of charged impurity scattering in the 
low density systems, which controls the physics of the 2D M-I-T. We believe that in the 
insulating phase ($n_s<n_c$) all the electrons are bound in pairs (or, some other small 
numbers) producing an insulating state whereas in the metallic phase ($n_s>n_c$), there 
are excess free electrons, which scatter from ions with bound pairs (or, some other 
small number) of electrons. We can qualitatively explain understand the density/temperature/magnetic field/electric field dependence of the 
observed resistivity both on the metallic and the insulating sides of the transition 
using the FELPI scenario, and also the fact that, at the low electron densities involved in 
the problem, the 2D M-I-T is experimentally close to a quantum-classical crossover 
regime. Our theory naturally provides an explanation for why $n_c\sim 10^{11} cm^{-2}$ 
in Si MOSFETs -- it is simply related to the ionic charge density being around a few 
times $10^{10} cm^{-2}$, provided high-mobility MOSFETs are used and the 
electron-electron interaction dominates over the kinetic energy part of the electrons.
Quantum phase transition 
considerations, which have created 
considerable excitement in the literature, turn out to be irrelevant to (and beyond the 
scope of) our proposed mechanism. Our theory makes precise predictions and is 
falsifiable -- for example, any experimental observation of a 2D M-I-T in Si-MOSFETs in 
the surface roughness scattering dominated regime {\it i.e.} $n_c>n_{max}$ will prove our theory wrong. Another 
example is our prediction that any lowering of the random ionic charge density will 
automatically push $n_c$ lower. While many theoretical details of our proposed FELPI 
scenario remain to be worked out, our proposal has the merit of being precise and 
concrete, which should enable one to calculate its narrow consequences to test the 
quantitative validity of FELPI for the 2D M-I-T problem.

\subsection*{Acknowledgments}
We like to thank S.V. Kravchenko for 
stimulating discussions and correspondence about the experimental results. Our work is 
supported by the NSF (both at QUEST and ITP, UCSB and at UMCP) and the ONR (SDS).In 
addition TMK likes to thank the Netherlands-America Committee for Educational 
Exchange for a Fulbright grant to partially support his stay at UCSB.     


\end{document}